\input{aipcheck}

\documentclass[
    ,final            
  ]
  {aipproc}

\layoutstyle{6x9}

\newcommand{\beqa}{\begin{eqnarray}}
\newcommand{\eeqa}{\end{eqnarray}}
\newcommand{\beq}{\begin{equation}}
\newcommand{\eeq}{\end{equation}}
\newcommand{\p}{\partial}
\newcommand{\too}{\rightarrow}

\begin{document}

\title{On the Equation of State of the Gluon Plasma\footnote{Quark Confinement and Hadron Spectrum VII, 2-7 September, 2006, Ponta Delgada, Azores, Portugal}}


\author{Daniel Zwanziger}{
  address={New York University, New York, NY 10003, USA}
}



\begin{abstract}
 We consider a local, renormalizable, BRST-invariant action for QCD in Coulomb gauge that contains auxiliary bose and fermi ghost fields.  It possess a non-perturbative vacuum that spontaneously breaks BRST-invariance.  The vacuum condition leads to a gap equation that introduces a mass scale.  Calculations are done to one-loop order in a perturbative expansion about this vacuum.  They are free of the finite-$T$ infrared divergences found by Lind\'{e} and which occur in the order $g^6$ corrections to the  Stefan-Boltzmann equation of state.  We obtain a finite result for these corrections.  
\end{abstract}

\maketitle


\section{Introduction}

	In 1980 Lind\'{e}~\cite{Linde:1980} showed that standard finite-temperature perturbation theory suffers from infrared divergences.  Since then no solution has been found, although the infrared divergences may be avoided by introducing a magnetic mass for the gluon in an ad hoc manner.  Some time ago, it was suggested that these divergences arise because the suppression of infrared gluons by the proximity of the Gribov horizon in infrared directions is neglected in standard perturbation theory~\cite{Zahed:1999}.  We present an approach in which is based on a local action that possesses a non-perturbative vacuum.  It appears that the infrared divergences found by Lind\'{e} do not arise in this approach because the free propagator, in this vacuum, of 3-dimensionally transverse, would-be physical gluons is strongly suppressed in the infrared.

Starting from the local action, we shall derive a gap equation that determines the non-perturbative vacuum.  We shall solve the gap equation at high temperature to obtain the leading non-perturbative correction to the Stefan-Boltzmann equation of state of the gluon plasma.  Additional results in the present approach, including propagators at the one-loop level, relation to the cut-off at the Gribov horizon, issues of principle concerning unitarity and Lorentz invariance, relation to other approaches, accord of the present approach with numerical studies, and references, may be found in \cite{Zwanziger:2006}.

\section{Local BRST-invariant action}

	We shall be interested in pure SU(N) gauge theory at temperature $T$.  Finite $T$ is described by a Euclidean action which, for pure SU(N) gauge theory, is of the form  
\beq
\label{lagdensity}
S = \int d^Dx \ {\cal L}; \ \ \ \ \ \ \ \  
{\cal L} = {\cal L}_{YM} + s \xi.
\eeq	
where
${\cal L}_{YM} = {1 \over 4} \ F_{\mu\nu}^2$,
and
$F_{\mu \nu} = \p_\mu A_\nu - \p_\nu A_\mu + g A_\mu \times A_\nu$.
Here $g$ is the coupling constant, and we use the notation for the Lie bracket 
$(A \times B)^a \equiv f^{abc} A^b B^c$,
where $f^{abc}$ are the fully anti-symmetric structure constants of the SU(N) group.  The color index is taken in the adjoint representation, $a = 1,..., N^2 -1$.  We shall generally  
suppress the color index, and leave summation over it implicit.  Configurations are periodic in~$x_0$, 
$A_\mu(x_i, x_0) = A_\mu(x_i, x_0 + \beta)$, 
with period $\beta = 1/T$, where $T$ is the temperature.  The integral over $x_0$ always extends over one cycle, 
$\int dx_0 \equiv \int_0^\beta dx_0$.  We are in $D$ Euclidean dimensions.   Lower case Latin indices take values, $i = 1, 2, ..., D-1$, while lower case Greek indices take values $\mu = 0, 1, ..., D-1$.

	In the BRST formulation there are, in addition to $A_\mu$, a pair of Faddeev-Popov ghost fields $c$ and $\bar c$ and a Lagrange multiplier field, $b$, on which the BRST operator acts according to
\beq
\label{BRST1}
s A_\mu = D_\mu c; \ \ \ \ \ \ \  \ \ \ \ \  
sc = - (g/2) c \times c; \ \ \ \ \ \ \ \ \ \ 
s \bar c = ib; \ \ \ \ \ \ \ \  \ \ \ \ \ \ \    s b = 0. 
\eeq	
It is nil-potent, $s^2 = 0$.  Here $D_\mu$ is the gauge-covariant derivative in the adjoint representation, $D_\mu c \equiv \p_\mu c + g A_\mu \times c$.

	The choice of the gauge-fixing density $\xi$ in
(\ref{lagdensity}) is the choice of gauge.  Physics is independent of $\xi$, provided that it results in a well-defined calculational scheme.  For finite $T$, this is a crucial proviso, because the standard gauge choice leads to infrared divergences \cite{Linde:1980}.  The standard Coulomb gauge is defined by $\xi = \xi_{coul} =  \p_i \bar c A_i$, with 
$s \xi = i \p_i b A_i - \p_i \bar c D_i c$.  The Lagrange-multiplier field $b$ imposes the Coulomb gauge condition $\p_i A_i = 0$.  To avoid the infrared divergences of the standard gauge choice we introduce an additional $s$-exact term $s\xi_{aux}$ that involves a quartet of auxiliary ghost fields on which $s$ acts trivially
\beq
\label{BRST2}
s \phi_\mu^{ab} = \omega_\mu^{ab}; 
\ \ \ \ \ \ \ \  s \omega_\mu^{ab} = 0;
\ \ \ \ \ \ \ \ 
s \bar\omega_\mu^{ab} = \bar\phi_\mu^{ab}; 
\ \ \ \ \ \ \ \  s \bar\phi_\mu^{ab} = 0.
\eeq	
The fields $\phi_\mu^{ab}$ and $\bar\phi_\mu^{ab}$ are a pair of bose ghosts, while $\omega_\mu^{ab}$ and $\bar\omega_\mu^{ab}$ are fermi ghost and anti-ghost.  The indices $a$ and $b$ label components in the adjoint representation of the global gauge group, $a, b =  1,..., N^2 -1$, and $\mu$ is a Lorentz index.  We take for the gauge-fixing term 
\beq
\label{gaugedensity}
\xi = \xi_{coul} + \xi_{aux} =  \p_i \bar c^a A_i^a 
+ \p_i \bar\omega_\mu^{ab} (D_i \phi_\mu)^{ab},
\eeq
where we stipulate that the gauge covariant derivative acts on the {\it first} color index only,
$(D_i \phi_\mu)^{ab} = \p_i \phi_\mu^{ab} + g f^{acd}A_i^c\phi_\mu^{db}$  etc.

\section{Non-perturbative vacuum}

\subsection{Maggiore-Schaden shift}

	We make a change of variable whereby the bose ghosts are translated by a $c$-number term linear in the spatial coordinate $x_\mu$ \cite{Maggiore:1994},
\beq
\label{MSshift}
\phi_\mu^{ab}(x) =  \varphi_\mu^{ab}(x) 
-  \gamma^{1/2} \  \delta^{ab} \  x_\mu; 
\ \ \ \ \ \ \ \ \ \ \ \ 
\bar\phi_\mu^{ab}(x) =  \bar\varphi_\mu^{ab}(x) 
+ \gamma^{1/2} \  \delta^{ab} \  x_\mu.
\eeq	
Note that $\phi$ and $\varphi$ designate different fields.  Here $\gamma$ is a parameter with dimensions of (mass)$^4$ that will be determined by the condition
${\p W \over \p \gamma} = 0$, where $W$ is the free energy.  We also translate $b$ and $\bar c$ by compensating terms,
\beq
\bar c^d = \bar c^{\star d} 
+ \gamma^{1/2} g f^{adb} x_\mu \bar\omega_\mu^{ab};
\ \ \ \ \ \ \ \ \ \ \ \ \ \ \ \ 
b^c = b^{\star c} 
-i \gamma^{1/2} g f^{acb} x_\mu \bar\varphi_\mu^{ab},
\eeq
which are chosen to cancel explicit $x$-dependence in the new action.  The BRST operator $s$ acts on the new variables according to
\beq
\label{BRST3}
s\varphi_\mu^{ab} = \omega_\mu^{ab};
\ \ \ \ \ \ \ \ 
s\bar\omega_\mu^{ab} = \bar\varphi_\mu^{ab}
+ \gamma^{1/2} \  \delta^{ab} \  x_\mu;
\ \ \ \ \ \ \ \ \ \ 
s \bar c^{\star d} = i b^{\star d}; \ \ \ \ \ \   s b^{\star d} = 0.
\eeq

	Despite the $x$-dependent shift, remarkably, neither the gauge-density (\ref{gaugedensity}) nor the Lagrangian density acquire any explicit $x$-dependence when expressed in terms of the new variables,
\beq
\xi = \p_i \bar c^{\star a} A_i^a
+ \p_i \bar\omega_\mu^{ab} (D_i \varphi_\mu)^{ab}
 -  \gamma^{1/2} \ (D_i \bar\omega_i)^{aa}.
\eeq
As before, it is understood that the gauge-covariant derivative acts on the first index only, 
$(D_i \bar\omega_\mu)^{ab} = \p_i \bar\omega_\mu^{ab}
+ g f^{acd} A_i^c \bar\omega_\mu^{db}$.  After the shift, the Lagrangian density is given by 
${\cal L} = {\cal L}_{YM} + s\xi$, where
\beqa
\label{Ldensity}
s\xi & = & i \p_i b^{\star a} A_i^a 
- \p_i \bar c^{\star a} (D_i c)^a
+ \p_i \bar\varphi_\mu^{ab} (D_i \varphi_\mu)^{ab}
 - \p_i \bar\omega_\mu^{ab} 
[ \ (D_i \omega_\mu)^{ab} + (g D_i c \times \varphi_\mu)^{ab} \ ]
\nonumber   \\  &&
+  \gamma^{1/2} \ (D_i \varphi_i)^{aa}
 -   \gamma^{1/2} \ 
 [ \ (D_i \bar\varphi_i)^{aa} + (gD_i c \times \bar\omega_i)^{aa} \ ]  
-  \gamma \ (N^2 -1) (D-1),
\eeqa
and $(D_i c \times \varphi_\mu)^{ab} \equiv f^{acd}(D_i c)^c \varphi^{db}$ acts on the first color index, etc.  For purposes of expansion in powers of $g$ we shall change indepdendent parameter from $\gamma$ to $m$ according to 
$\gamma^{1/2} \equiv {m^2 \over (2N)^{1/2}g}$,
where $m$ has dimensions of mass, and $m$ is taken to be of order $g^0$.

\subsection{Gap equation}

	Henceforth we shall be concerned with the action $S$, regarded as a function of the new fields.  The partition function is given by $Z = \int d\Phi \exp(-S)$,
where $\Phi \equiv (A_\mu, c, \bar c, b, \varphi, \bar\varphi, \omega, \bar\omega)$ is the set of all (new) fields, and $d\Phi$ represents integration over them.  For simplicity we have written $\bar c$ and 
$b$ instead of $\bar{c}^\star$  and $b^\star$.  The field $\varphi_\mu^{ab}$ is real while $\bar\varphi_\mu^{ab}$ is pure imaginary.  The classical vacuum occurs where all these fields vanish 
$\Phi \equiv (A_\mu, c, \bar c, b, \varphi, \bar\varphi, \omega, \bar\omega) = 0$.    
Finally the value of $\gamma$ is determined by the condition that the free energy $W = \ln Z$ be stationary,
${ \p W \over \p  \gamma^{1/2} } = 0$,
or
$\Big\langle { \p S \over \p  \gamma^{1/2} } \Big\rangle = 0$.
There is a non-perturbative vacuum if this equation has a solution with $\gamma \neq 0$.  We do not require that $W$ be a maximum because there are unphysical fields present.  The last equation reads
\beq
\label{horizoncondb}
\langle  D_i (\varphi_i - \bar\varphi_i)^{aa} 
 - (gD_i c \times \bar\omega_i)^{aa} \rangle 
=  2  \gamma^{1/2}  (N^2-1)(D-1).
\eeq
The second term vanishes,
$\langle (gD_i c \times \bar\omega_i)^{aa} \rangle = 0$,
because there is no $\bar c \omega$ term in the action.  Moreover the new Lagrangian density (\ref{Ldensity}) is invariant under space-time translation of the new fields, $\Phi(x) \too \Phi(x + a)$, and the vacuum just found, at $\Phi = 0$, is also.  Translation invariance implies that the terms $\p_i \varphi$ and $\p_i \bar\varphi$ do not contribute to (\ref{horizoncondb}), and we obtain 
\beq 
\label{horizoncond}
 { 1  \over (2N)^{1/2} } \ \langle f^{abc} A_i^b 
 (\varphi - \bar\varphi)_i^{ca} \rangle
 = {m^2 \over N g^2} \ (D-1)(N^2 -1). 
\eeq
This gap equation determines $m = m(g, T)$.  Invariance under scale transformation is spontaneously broken for $m \neq 0$.

\subsection{Spontaneous breaking of BRST symmetry}

	The quantity $m^2$ is the analog of the vacuum expectation-value $v$ of the Higgs field $\Phi$ that appears in spontaneous symmetry breaking of global gauge invariance.  For in the Higgs mechanism one makes the translation
$\Phi^a = \Phi^{\star a} + v \delta_3^a$, and the vacuum expectation-value $v$ is determined by the condition that the free-energy be stationary with respect to $v$, 
${ \p W \over \p v } = 0$.  Similarly, $m^2$ is determined by the condition ${\p W \over \p m} = 0$.  In the present case, BRST invariance is spontaneously broken rather than global gauge invariance, because the expectation-value of $s$-exact quantities is non-zero, for example,
\beq
\label{breakexp}
\langle \ s \bar\omega_\mu^{ab} \ \rangle = 
\langle \  \bar\varphi_\mu^{ab} 
+ \gamma^{1/2} \delta^{ab} \ x_\mu \   \rangle
=  \gamma^{1/2} \delta^{ab} \ x_\mu \neq 0.
\eeq 

	As in the Higgs case, the spontaneously broken theory inherits renormalizability from the unbroken theory.  But (\ref{breakexp}) shows that we cannot identify observables with equivalence classes of $s$-invariant objects, modulo $s$-exact quantities, as in the standard BRST approach.  However, as shown in \cite{Zwanziger:2006}, the present method is formally equivalent to the canonical formulation of Coulomb gauge, with a cut-off at the Gribov horizon.  This allows us to identify observables, such as the energy-momentum tensor $T_{\mu \nu}$, with the corresponding quantities in the canonical formulation so, for example,
$T_{\mu \nu} = F_{\mu \lambda} F_\nu^\lambda - {1 \over 4} g_{\mu \nu} F_{\kappa \lambda} F^{\kappa \lambda}$.

\section{Free propagators}

	We now develop a perturbative expansion about the new, non-perturbative, vacuum.  For this purpose we treat $m$ as an independent parameter of order $g^0$, and calculate perturbatively all one-particle irreducible graphs, including the gap equation, to a given order in $g$.  Then $m = m(g, T)$ is determined by solving the the gap equation (\ref{horizoncond}) non-perturbatively.  
	
	The first step is to expand the action in powers of~$g$, $S = S_{-2} + S_0 + ... \ $.  
The leading term is of order $g^{-2}$,
\beq
\label{minus2}
S_{-2} \equiv - {m^4 \over 2Ng^2} \ (N^2 -1) (D-1)L^3 \beta,
\eeq	
where the spatial quantization volume is $V = L^3$, and the time extent $\beta = T^{-1}$.  Although $S_{-2}$ is independent of the fields, and does not contribute to the propagators,  it should not be ignored because, when the gap equation is solved for $m = m(T)$, it gives a $T$-dependent contribution to the free energy.  The terms in the action of order $g^0$ are all quadratic in the fields
\beqa
\label{S0}
S_0  = & 
\int d^Dx & \Big[ {1 \over 4} 
\p_\mu A_\nu^a - \p_\nu A_\mu^a)^2 
+ i \p_i b^a A_i^a - \p_i \bar c^a \p_i c^a
\nonumber   \\   &&
+ \p_i \bar\varphi_\mu^{ab} \p_i \varphi_\mu^{ab} 
- \p_i \bar\omega_\mu^{ab} \p_i \omega_\mu^{ab}
+ { m^2 \over (2N)^{1/2} } \ f^{abc} A_i^b 
 (\varphi_i - \bar\varphi_i)^{ca}
 \Big],
\eeqa	
 and determine the ``free" propagators.  The term with coefficient $m^2$ causes a mixing of the zero-order transverse gluon and bose-ghost propagators.  

	To calculate the free propagators, we define the field, $\psi_j^b \equiv { i \over (2 N)^{1/2} }   f^{abc}
 (\varphi_j^{ca} - \bar\varphi_j^{ca})$, that mixes with $A_i^b$.  The orthogonal component ${ 1 \over (2 N)^{1/2} }   f^{abc}
 (\varphi_j^{ca} + \bar\varphi_j^{ca})$ and other components of $\varphi$ and $\bar\varphi$ do not mix with $A_i$.  The free propagators are given by
\beq
\label{glueprop}
D_{A_iA_j}({\bf k}, k_0)  =   
{ P_{ij} \ {\bf k}^2 \over \Delta }; \ \ \ \ \ \ 
D_{A_i \psi_j}({\bf k}, k_0) =   
{P_{ij} \ im^2 \over \Delta }; \ \ \ \ \ \ \ 
D_{\psi_i \psi_j}({\bf k}, k_0) =   
{ P_{ij} \ ({\bf k}^2 + k_0^2) \over \Delta }.
\eeq 
where $\Delta \equiv  ({\bf k}^2 + k_0^2){\bf k}^2 + m^4$,
and $P_{ij} \equiv \delta_{ij} - \hat k_i \hat k_j$ is the transverse projector.  Here $k_0 = 2\pi n/\beta$ are the Matsubara frequencies, where $n$ is any integer, and we have suppressed the trivial color factor $\delta^{bc}$.  In terms of the variable $\psi_j^a$, the gap equation (\ref{horizoncond}) reads 
\beq 
\label{horizonconda}
 -i \ \langle A_j^c(0) \psi_j^c(0) \rangle
 = {m^2 \over N g^2} \ (D-1)(N^2 -1). 
\eeq

\section{Gap equation in one-loop approximation}

When the left-hand side of the gap equation is evaluated to zeroth order in $g$, using the mixed propagator~(\ref{glueprop}), it reads
\beq
\label{gaphorizon}
 \int { d^{D-1}k \over (2\pi)^{D-1} } \ 
T\sum_{k_0} 
 { D-2 \over ({\bf k}^2 + k_0^2){\bf k}^2 + m^4 } = {D-1 \over Ng^2 },
\eeq
where we used $P_{ii}({\bf k}) = D-2$.  The sum over Matsubara frequencies yields
\beq
\label{gapeq}
\int { d^{D-1}k \over  (2 \pi)^{D-1} } \ { (D - 2) \over 2 {\bf k}^2 E }
 \ \Big(1 + { 2 \over \exp(\beta E) -1 }\Big) = { D - 1 \over N g^2 },
\eeq
which holds in $D$ Euclidean space-time dimensions.  The first term in parentheses gives the result at $T = 0$, and  the second term is a Planck-type finite-temperature correction.  For $D < 4$ the integral is convergent.  We take the limit $D \too 4$.  The first term in parentheses has the limiting form 
\beq
\int { d^{D-1}k \over  (2 \pi)^{D-1} } \ { (D - 2) \over 2 {\bf k}^2 E }
\too { 1 \over 4 \pi ^2 } \Big[ {1 \over \epsilon } 
+ \ln \Big(  {\mu^2 \over m^2 } \Big) \Big]
\eeq
where $\epsilon \equiv (4 - D)/2$, and $\mu = \mu(T)$ is, in general, a temperature-dependent renormalization mass.  There is a pole at $D = 4$.  We subtract the pole term, and the gap equation reads
\beq
\label{gapeq2}
 { 1 \over 2  }   \ln \Big(  {\mu \over m } \Big) 
+  \int_0^\infty { dx \over u }
\ { 1 \over \exp(m \beta u) -1 }  = { 3 \pi^2 \over N g^2(\mu) },
\eeq
where $u \equiv (x^2 + {1 \over x^2 } )^{1/2}$.  
	
	We now specialize to high temperature $T$, and take the renormalization mass to be $\mu = T$.  This simplifies the high-T case because of asymptotic freedom.  The running coupling $g(T)$ is small, and is given approximately by
${1 \over g^2(T)} = {11 \ N \over 24 \ \pi^2} 
\  \ln \Big( {2 \pi T \over \Lambda_{{\overline{MS}}} } \Big)$, where $\Lambda_{{\overline{MS}}}$ is the ${\overline{MS}}$ physical mass scale.  At high temperature, the first term in (\ref{gapeq2}) may be neglected, and the second term has the limit
\beq
\int_0^\infty { dx \over u } \ { 1 \over \exp(\beta m u) -1 }
\too { 1 \over \beta m } \int_0^\infty dx \  { x^2 \over x^4 + 1 } 
= { \pi T \over 2^{3/2} \ m },
\eeq
so the gap equation at high $T$ simplifies to
${ \pi T \over 2^{3/2} \ m } = { 3 \ \pi^2 \over N \ g^2(T) }$, with solution,
\beq
\label{mofT}
m(T, g) = { N   \over 2^{3/2}  \ 3 \  \pi } \ g^2(T) \ T
 \ \ \ \ \ \ \ \ \    T \too \infty.
\eeq
Thus, in the high-temperature limit, $m(T)$ is proportional to the magnetic mass $g^2(T) T$.

\section{Free energy}

	To order $g^0$, the free energy $W = \ln Z$ is given by 
\beq
\exp W = \int d \Phi \ \exp( - S_{-2} - S_0 ),
\eeq
where $S_{-2}$, given in (\ref{minus2}), is field-independent and of order $g^{-2}$, while $S_0$, given in (\ref{S0}), is quadratic in the fields.  We obtain
$W = W_{-2} + W_0$
where, for $D -1 = 3$,
$W_{-2} = - S_{-2} =  {3m^4 \over 2Ng^2} \ (N^2 -1) L^3 \beta$, and
$\exp W_0 =  \int d \Phi \exp( - S_0 )$.  The evaluation of the free energy $W_0$ for the quadratic action $S_0$ is straightforward, keeping in mind the $A$-$\varphi$ and $A$-$\bar\varphi$ mixing, with the result   
\beq
W_0 =  { (N^2 -1)V \beta \over 3 \pi^2 } 
\int_0^\infty dk \ { (k^4 -m^4) \over E \  [\exp(E \beta ) -1] },
\eeq
where $E = E(k) = (k^2 + { m^4 \over k^2 } )^{1/2}$.  We add the term $W_{-2}$ and obtain to order $g^0$ the free energy per unit volume, $w = W/V$,
\beq
\label{freeenergy}
w = (N^2 -1) \ \beta \ \Big( { 3 \ m^4  \over 2 N g^2 }
+ { 1 \over 3 \pi^2 } 
\int_0^\infty dk \ { (k^4 -m^4) \over E \  [\exp(E \beta ) -1] } \Big).
\eeq

\section{Equation of state at high temperature}

	We now evaluate $w$ in the high-temperature limit, where $g(T)$ is small, and our expansion should be reliable.  We insert the solution, 
$m =  { N \ g^2(T) \ T \over  2^{3/2} \ 3 \ \pi }$, of the gap equation and obtain	
\beq
\label{freeenergy}
w = (N^2 -1) \ \Big( {  N^3 \ g^6(T)   \over 2^7 \ 3^3 \  \pi^4  }
+ { 1 \over 3 \pi^2 } K(\eta) \Big) \ T^3 ,
\eeq
where
$K(\eta) \equiv \int_0^\infty { dy \ (y^4 - \eta) \over u \  (\exp u  -1) }$; 	
$u \equiv (y^2 + {\eta \over y^2})^{1/2}$,
and
$\eta \equiv \Big({m \over T}\Big)^4 
= \Big( { N \ g^2(T) \over  2^{3/2} \ 3 \ \pi } \Big)^4$
is a small parameter.  With neglect of higher order terms, one obtains $K(\eta) = { \pi^4 \over 15 } - { \pi \ \eta^{3/4} \over 2^{1/2} }$, which gives
\beq
w = (N^2 -1) \ \Big( {\pi^2 \over 45 }
- { N^3   \over 10,\!368 \ \pi^4  }  \ g^6(T)  \Big) \ T^3.
\eeq
This is the leading correction to the Stefan-Boltzmann limit from the non-perturbative vacuum.  The equation of state of the gluon plasma follows from the thermodynamic formulas for the energy per unit volume and pressure,
$e = - { \p w \over \p \beta }$; $ p = { w \over \beta }$,
and entropy per unit volume, $s = {e + p \over T}$.  To calculate the energy density, we use $-\beta {\p g \over \p \beta} = T {\p g \over \p T} = \beta$-function $=  O(g^3)$, which is of higher order.  We thus obtain for the energy density and pressure at high temperature, 
$e = 3p = w(T) T$, and $s = { 4 \over 3 }w(T)$.

	We have obtained the leading contribution that comes from the non-perturbative vacuum.  Numerically it is a small correction, whereas the correction of order $g^6$ is divergent when calculated with the perturbative vacuum \cite{Linde:1980}.  To this must be added the perturbative contributions, including resumations, that have been calculated  at $m = 0$, and that are of lower order in $g$ \cite{Kapusta:1989}.  Since standard, resummed perturbation theory diverges at order $g^6$, which is precisely the order of the correction we have found, the result obtained here is consistent with standard perturbative calculations.





\begin{theacknowledgments}
I recall with pleasure stimulating conversations about this work with Reinhard Alkofer, Laurent Baulieu, David Dudal, Andrei Gruzinov, Atsushi Nakamura, Robert Pisarski, Alexander Rutenburg, Martin Schaden, and Silvio Sorella. 
\end{theacknowledgments}









\end{document}